\documentclass{iopart}
\input epsf
\begin{document}

\title[Wehrl information entropy and phase distributions]
{Wehrl information entropy and phase distributions
of~Schr\"odinger cat and cat-like states}

\author{
A Miranowicz,$^{1,2}$ J Bajer,$^3$ M R B Wahiddin$^4$ and N
Imoto$^1$}

\address{$^1$
CREST Research Team for Interacting Carrier
Electronics, School of Advanced Sciences, The Graduate
University for Advanced Studies (SOKEN), Hayama, Kanagawa
240-0193, Japan}

\address{$^2$
Nonlinear Optics Division, Institute of Physics,
Adam Mickiewicz University, 61-614 Pozna\'n, Poland}

\address{$^3$
Department of Optics, Palack\'{y} University,
17.~listopadu 50, 772~00 Olomouc, Czech Republic}

\address{$^4$
Institute of Mathematical Sciences, University of Malaya, 50603 Kuala
Lumpur, Malaysia}

\date{\today}

\begin{abstract}
The Wehrl information entropy and its phase density, the so-called
Wehrl phase distribution, are applied to describe Schr\"odinger
cat and cat-like (kitten) states.  The advantages of the Wehrl
phase distribution over the Wehrl entropy in a description of the
superposition principle are presented. The entropic measures are
compared with a conventional phase distribution from the Husimi
$Q$-function. Compact-form formulae for the entropic measures are
found for superpositions of well-separated states. Examples of
Schr\"odinger cats (including even, odd and Yurke-Stoler coherent
states), as well as the cat-like states generated in Kerr medium
are analyzed in detail. It is shown that, in contrast to the Wehrl
entropy, the Wehrl phase distribution properly distinguishes
between different superpositions of unequally-weighted states in
respect to their number and phase-space configuration.

%\vspace{2mm}{PACS numbers: 42.50.Dv, 42.50.Md}
\end{abstract}

\vspace{5mm}

%%%%%%%%%%%%%%%%%%%%%%%%%%%%%%%%%%%%%%%%%%%%%%%%%%%%%%%%%%%%%%%%%%%%%%%%%%%%

\section{Introduction}

Schr\"{o}dinger cats or cat-like states (kittens) are the striking
manifestations of the superposition principle at boundary between
the quantum and classical regimes~\cite{Schr35}. Especially since
the 1980s, the Schr\"{o}dinger cats have attracted much interest
in quantum and atom optics or quantum computing by allowing
controlled studies of quantum measurement, quantum entanglement
and decoherence. Further interest has recently been triggered by
first experimental demonstrations of Schr\"{o}dinger cats on
mesoscopic~\cite{Monr96} and also macroscopic~\cite{Frie00}
scales.

We will analyze simple prototypes of the Schr\"{o}dinger cat and
cat-like states in entropic and phase descriptions. The Wehrl
classical information entropy is defined to be~\cite{Wehr78}:
%---------------------------------------------------------------------------
\begin{eqnarray}
S_{\rm w}\equiv -\int Q(\alpha )\ln Q(\alpha ){\rm d}^{2}\alpha
\label{N01}
\end{eqnarray}
where $Q(\alpha)$ is the Husimi function given by $Q(\alpha )=\pi
^{-1}\langle \alpha |\widehat{\rho }|\alpha \rangle$ as the
coherent-state representation of density matrix $\widehat{\rho}$.
The Wehrl classical information entropy, also
\linebreak
%%%%%%%%%%%%%%%%%%%%%%%%%%%%%%%%%%%%%%%%%%%%%%%%%%%%%%%%%%%%%%%%%%%%%%%%%%%%
%TABLE 1.
\begin{table}
\caption{ Are the Wehrl information entropy, its density or Husimi
PD good measures of the properties of classical and quantum
fields? }
\begin{center}
\begin{tabular}{l l l l l}
\hline
No. & Measures  & Wehrl   & Husimi & Wehrl \\
    &           & entropy & PD    & PD     \\
\hline
1. & discrimination of fields with random phase & yes  & no & yes \\
%-------------------------------------------------------------
2. & photon-number uncertainty  & yes & no & yes \\
%-------------------------------------------------------------
3. & discrimination of coherent states
with different intensity & no  & yes & yes \\
%--------------------------------------------------------------
4. & phase locking     & no  & yes & yes \\
%-------------------------------------------------------------
5. & phase bifurcation & no  & yes & yes \\
%-------------------------------------------------------------
6. & Schr\"odinger cats
with different weights & no   & yes & yes \\
%-------------------------------------------------------------
7. & Schr\"odinger cat-like states with different weights & no &
yes &
yes \\
%-------------------------------------------------------------
\hline
\end{tabular}
\end{center}
\end{table}
\noindent referred to as the Shannon information of the Husimi
$Q$-function, can be related to the von Neumann quantum entropy in
different approaches \cite{Wehr79,Peri86}, in particular in
relation to a phase-space measurement \cite{Buze95a}. It has been
demonstrated that the Wehrl entropy is a useful measure of various
quantum-field properties, including quantum
noise~\cite{Peri86}--\cite{Wats96}, decoherence
\cite{Ande93,Orlo95}, quantum interference \cite{Buze95b},
ionization \cite{Wats96}, or squeezing~\cite{Keit92,Lee88,Orlo93}.
Moreover, it has been shown that the Wehrl entropy gives a clear
signature of the formation of the Schr\"{o}dinger cat and cat-like
states~\cite{Vacc95,Jex94} and a signature of splitting of the
$Q$-function~\cite{Orlo95}. Here, we will show explicitly that the
Schr\"{o}dinger cat and cat-like states are, in general, {\em not}
uniquely described by the conventional Wehrl entropy. Thus, for
better description of Schr\"{o}dinger  cat and cat-like states, we
apply another entropic measure -- the so-called Wehrl phase
distribution (Wehrl PD), defined to be the phase density of the
Wehrl entropy \cite{Mira00}:
%---------------------------------------------------------------------------
\begin{eqnarray}
S_{\theta }\equiv -\int Q(\alpha )\ln Q(\alpha )|\alpha |{\rm
d}|\alpha| \label{N02}
\end{eqnarray}
where $\theta ={\rm Arg}\alpha$. The Wehrl PD is simply related to
the Wehrl entropy via integration, i.e., $S_{\rm w}=\int S_{\theta
}{\rm d}\theta$. The Wehrl PD is formally similar to  a
conventional phase distribution from the Husimi $Q$-function
defined to be (see, e.g., \cite{Tana96})
%---------------------------------------------------------------------------
\begin{eqnarray}
P_{\theta }\equiv\int Q(\alpha )|\alpha |{\rm d}|\alpha |
\label{N03}
\end{eqnarray}
which is referred to as the Husimi phase distribution (Husimi PD).
The main physical advantage of the Wehrl PD over the conventional
PDs (including that of Husimi) lies in its information-theoretic
content. The Wehrl entropy, which is the area covered by the Wehrl
PD, is a measure of information that takes into account the
measuring apparatus (homodyne detection) used to obtain this
information. Various other advantages of the Wehrl PD over the
conventional PDs and the Wehrl entropy itself in a description of
quantum and classical optical states of light were demonstrated in
Ref. \cite{Mira00}. Several examples of applications of the Wehrl
PD are listed in table 1. Example 1 is a consequence of the
Wehrl-entropy sensitivity in discriminating different fields with
random phase. By contrast, the conventional PDs are the same for
arbitrary random-phase fields. Example 2 comes from the fact that
the Werhl entropy is directly related to the phase-space
measurement \cite{Buze95a} thus in particular carrying information
about the photon-number uncertainties. On the other hand, the
conventional PDs do {\em not} measure photon-number properties.
Examples 3--5, discussed in detail in Ref. \cite{Mira00},
%%%%%%%%%%%%%%%%%%%%%%%%%%%%%%%%%%%%%%%%%%%%%%%%%%%%%%%%%%%%%%%%%%%%%%%%%%%%
%figure 1.
\linebreak
\begin{figure}
\vspace*{0cm} \hspace*{2.5cm} \epsfxsize=10.5cm \epsfbox{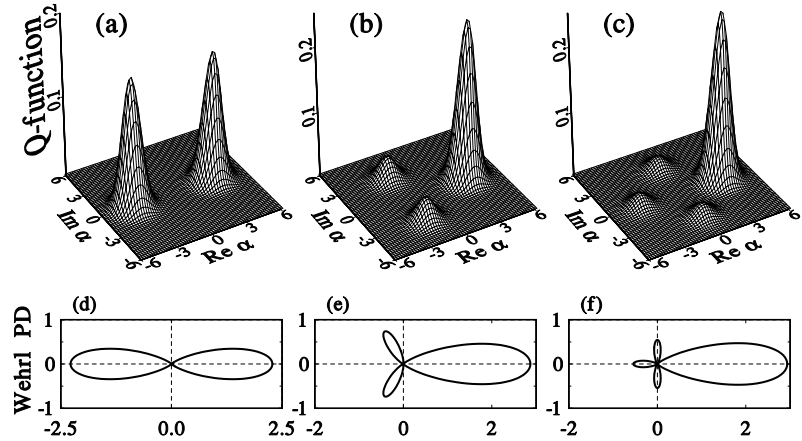}
\vspace*{0cm} \caption{Equientropic Schr\"odinger cat and cat-like
superpositions of $N=2,3,4$ states. Husimi $Q$-functions (a,b,c)
and the corresponding polar plots of Wehrl phase distributions
(d,e,f) for well-separated ($N<N_{\max}(\alpha_0=\sqrt{12})=7$)
superposition states described by the same Wehrl entropy. }
\end{figure}
\noindent show advantages of both the Wehrl and conventional PDs
over the Wehrl entropy. In the next sections we will compare
descriptions of the Schr\"{o}dinger cat and cat-like states in
terms of the Wehrl entropy, and the Wehrl and Husimi PDs. Our
analytical and numerical results are briefly summarized in table 1
by examples 6 and 7.

%%%%%%%%%%%%%%%%%%%%%%%%%%%%%%%%%%%%%%%%%%%%%%%%%%%%%%%%%%%%%%%%%%%%%%%%
\section{Equientropic cat and cat-like states}

First, we will show that the conventional Wehrl entropy does {\em
not} uniquely distinguish between different superpositions of
unequally-weighted states. Let us analyze in detail the following
superposition of $N$ coherent states
%----------------------------------------------------------------------
\begin{eqnarray}
\hspace{-5mm} |\overline{\alpha}_0\rangle_N &=&{\cal C}_N\Big\{
\sqrt{1-(N-1)x_{N}}|\alpha _{0}\rangle +\sqrt{x_{N}}
\sum_{k=1}^{N-1}|\exp \left( {\rm i}k\frac{2\pi }{N}\right) \alpha
_{0}\rangle\Big\} \label{N04}
\end{eqnarray}
where ${\cal C}_N$ is the normalization constant and $x_{N+1}$ are
the roots of
%----------------------------------------------------------------------
\begin{eqnarray}
f_{N+1}(x)=2(1-Nx)^{(1-Nx)}x^{Nx}-1. \label{N05}
\end{eqnarray}
In particular equation (\ref{N05}) has the following roots
$x_{2}=\frac{1}{2}$, $x_{3}=0.11\cdots$, $x_{4}=0.063\cdots$. For
$N=2$, the state $|\overline{\alpha}_0\rangle_2$ is the even
coherent state, given by (\ref{N10}). We observe that the quantum
superpositions $|\overline{\alpha}_0\rangle_N$ of well-separated
coherent states have the same Wehrl entropy equal to
%----------------------------------------------------------------------
\begin{eqnarray}
\lim_{(|\alpha _{0}|/N)\rightarrow\infty} S_{\rm
w}(|\overline{\alpha}_0\rangle_N)=1+\ln 2\pi \label{N06}
\end{eqnarray}
for any finite $N>1$. The $Q$-functions for three different states
with well-separated $N=2$, 3 and 4 components are depicted in
figures 1(a)--(c). In the high-amplitude limit, the interference
terms in $Q$-function vanish and the normalization ${\cal C}_N$
becomes one. The Wehrl entropy for these states is equal to
$S_{\rm w}=1+\ln 2\pi-\epsilon_N$ with the corrections: (a)
$\epsilon_2 < 0.000002$, (b)  $\epsilon_3 < 0.000065$, and (c)
$\epsilon_4 < 0.0012$ for the intensity $|\alpha _{0}|^2=12$. As
comes from (\ref{N06}), these corrections can be made arbitrary
small by increasing $|\alpha _{0}|$ in comparison to $N$. In fact,
for any quantum superposition of macroscopically distinct states,
other superposition states can be found with the same Wehrl
entropy but distinct number of components. Thus, one can conclude
that the conventional Wehrl entropy is not sensitive enough in
discriminating cats from unequally-weighted cat-like states. The
main purpose of this paper is to apply a new entropic measure and
to show its advantages over the conventional Wehrl entropy in
describing macroscopically distinct superpositions of states.

%%%%%%%%%%%%%%%%%%%%%%%%%%%%%%%%%%%%%%%%%%%%%%%%%%%%%%%%%%%%%%%%%%%%%%%%
\section{New entropic description of cat and cat-like states}

A standard example of the Schr\"{o}dinger cat  is a superposition
of two coherent states $|\alpha _{0}\rangle $ and $|-\alpha
_{0}\rangle $ in the form (see, e.g.,~\cite{Buze95cat}):
%---------------------------------------------------------------------------
\begin{eqnarray}
|\alpha _{0},\gamma \rangle &=&{\cal N}_{\gamma }\left\{ |\alpha
_{0}\rangle +\exp ({\rm i}\gamma )|-\alpha _{0}\rangle \right\}
\label{N07}
\end{eqnarray}
with normalization
%---------------------------------------------------------------------------
\begin{eqnarray}
{\cal N}_{\gamma }=\left\{ 2\left[ 1+\cos \gamma \exp (-2|\alpha
_{0}|^{2})\right] \right\} ^{-1/2}.\label{N08}
\end{eqnarray}
For special choices of the superposition phase $\gamma $, the
state $|\alpha _{0},\gamma \rangle$ reduces to the well-known
Schr\"{o}dinger cats, including the Yurke-Stoler coherent state
for $\gamma =\pi /2$ \cite{Yurk86}:
%---------------------------------------------------------------------------
\begin{eqnarray}
|\alpha _{0}\rangle _{{\rm YS}}&=&|\alpha _{0},\pi /2\rangle
=\frac{1}{ \sqrt{2}}\left( |\alpha _{0}\rangle +{\rm i}|-\alpha
_{0}\rangle \right) \label{N09}
\end{eqnarray}
and the even ($\gamma =0$) and odd ($\gamma =\pi $) coherent
states~\cite{Peri91}:
%---------------------------------------------------------------------------
\begin{eqnarray}
\hspace{-9mm} |\alpha _{0},0\rangle &=& {\cal N}_{0}\left( |\alpha
_{0}\rangle +|-\alpha _{0}\rangle \right) = \frac{1}{\sqrt{\cosh
|\alpha _{0}|^{2}}}\sum_{n=0}^{\infty }\frac{\alpha
_{0}^{2n}}{\sqrt{(2n)!} }|2n\rangle , \label{N10}\\
\hspace{-9mm} |\alpha _{0},\pi \rangle &=& {\cal N}_{\pi }\left(
|\alpha _{0}\rangle -|-\alpha _{0}\rangle \right) =
\frac{1}{\sqrt{\sinh |\alpha _{0}|^{2}}}\sum_{n=0}^{\infty
}\frac{\alpha _{0}^{2n+1}}{ \sqrt{(2n+1)!}}|2n+1\rangle
\label{N11}
\end{eqnarray}
respectively. The Husimi function for the Schr\"{o}dinger cat
$|\alpha _{0},\gamma \rangle$ can be given in form of the sum
%---------------------------------------------------------------------------
\begin{eqnarray}
Q(\alpha )&=&{\cal N}_{\gamma }^{2}[Q_{1}(\alpha )+2Q_{12}(\alpha
)+Q_{2}(\alpha )] \label{N12}
\end{eqnarray}
of the coherent components ($k=1,2$)
%---------------------------------------------------------------------------
\begin{eqnarray}
Q_{k}(\alpha )&=&\frac{1}{\pi }\exp \left\{ -\left| \alpha
+(-1)^{k}\alpha _{0}\right| ^{2}\right\}
\label{N13}
\end{eqnarray}
and the interference term
%---------------------------------------------------------------------------
\begin{eqnarray}
Q_{12}(\alpha )&=&\frac{1}{\pi }\exp (-|\alpha |^{2}-|\alpha
_{0}|^{2})\cos [\gamma +2|\alpha |\,|\alpha _{0}|\sin (\theta
_{0}-\theta )] \label{N14}
\end{eqnarray}
where $\theta _{0}$ is the phase of $\alpha _{0}$. There is no
compact-form exact expression of the phase distributions for the
Schr\"{o}dinger cat defined by (\ref{N07}). The states analyzed in
our former work \cite{Mira00} are among a few examples, where the
phase distributions can be expressed analytically in a compact
form. Nevertheless, for well-separated ($|\alpha _{0}|\gg 1$)
coherent states $|\alpha _{0}\rangle $ and $|-\alpha _{0}\rangle$,
we find that the Wehrl PD can be approximated by
%---------------------------------------------------------------------------
\begin{eqnarray}
S_{\theta }\approx \frac{1}{2}\Big\{&&S^{\rm cs}_{\theta }(\alpha
_{0})+S^{\rm cs}_{\theta }(-\alpha _{0}) +\ln 2\big[P^{\rm
cs}_{\theta }(\alpha _{0})+P^{\rm cs}_{\theta }(-\alpha _{0}
)\big]\Big\} \label{N15}
\end{eqnarray}
in terms of the Werhl and Husimi phase distributions for coherent
states \cite{Mira00}:
%----------------------------------------------------------------------
\begin{eqnarray}
S^{\rm cs}_{\theta }(\alpha _{0}) &=&\frac{1}{2\pi}\, {\rm
e}^{X^{2}-X_{0}^{2}} \Big\{ {\rm e}^{-X^2}f_{2}+\sqrt{\pi
}X\,[1+{\rm erf}(X)]f_{1}\Big\}
\label{N16} \\
P^{\rm cs}_{\theta }(\alpha _{0}) &=&\frac{1}{2\pi}\, {\rm
e}^{X^{2}-X_{0}^{2}} \Big \{{\rm e}^{-X^2} +\sqrt{\pi }X\,[1+{\rm
erf}(X)]\Big\} \label{N17}
\end{eqnarray}
%%%%%%%%%%%%%%%%%%%%%%%%%%%%%%%%%%%%%%%%%%%%%%%%%%%%%%%%%%%%%%%%%%%%%%%%%%%%
%figure 2.
\linebreak
\begin{figure}
\vspace*{2mm} \hspace*{2.5cm} \epsfxsize=10.5cm \epsfbox{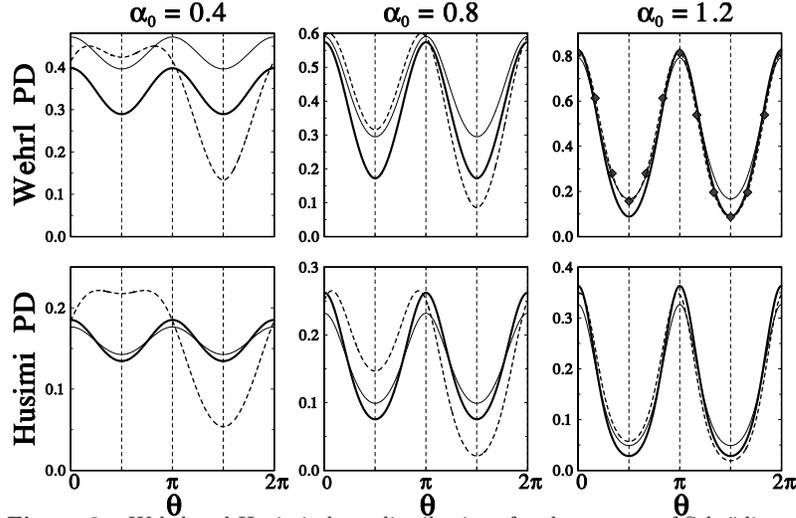}
\vspace*{-5mm} \caption{ Wehrl and Husimi phase distributions for
three types of Schr\"odinger cats: even (thick solid curves), odd
(thin solid curves), and Yurke-Stoler (dashed curves or those
with diamonds) coherent states for different values of the
coherent amplitude $\alpha_0$. } \label{fig2} \vspace*{-5mm}
\end{figure}
\noindent
respectively, where
%---------------------------------------------------------------------------
\begin{eqnarray}
f_{j}&=&X_{0}^{2}-X^{2}+\ln \pi +j/2, \nonumber\\
X&=&|\alpha _{0}|\cos (\theta -\theta _{0}) \label{N18}
\end{eqnarray}
and $X_{0}=X(\theta =\theta _{0})=|\alpha _{0}|$. Moreover, ${\rm
erf}(X)$ is the error function. As was discussed in Ref.
\cite{Mira00}, the Wehrl and Husimi PDs for coherent states differ
by the factors $f_{j}$ only. The $Q$-function and Wehrl PD for the
cat $|\alpha _{0},\gamma \rangle$  with well-separated components
($|\alpha _{0}| \gg 1$) are presented in figures 1(a) and 1(d),
respectively. In this case, the contribution of interference term
$Q_{12}(\alpha)$ is negligible. Thus, the $Q$-function and Wehrl
PD for $|\alpha _{0}|^2=12$ are practically independent of the
superposition coefficient $\gamma$, and the difference between the
exact Wehrl PD and its approximation, given by (\ref{N15}),
vanishes. This can be understood better by analyzing figure 2,
where the Wehrl and Husimi PDs are depicted for different values
of separation amplitude $\alpha _{0}$ for three superposition
parameters $\gamma$ corresponding to the Yurke-Stoler, even and
odd coherent states. It is apparent, both in the Wehrl and Husimi
PDs, that the differences among cats described by
(\ref{N09})--(\ref{N11}) diminish with increasing amplitude
$\alpha _{0}$ and would completely disappear even at $|\alpha
_{0}|=2.4$ on the scale of figure 2. It is also seen that the
maximum values of $P_{\theta}$ and $S_{\theta}$ depend on the
separation amplitude $|\alpha_{0}|$. However, only in case of the
Wehrl PD, the area under the curve is amplitude-dependent being an
indicator of the phase-space uncertainty. Despite of the formal
similarities, the Wehrl and Husimi PDs differ recognizably for
superposition of states, which are not well separated ($|\alpha
_{0}|<1$). For example, the Wehrl PD for even coherent state is
less than that for odd cat for all phases $\theta$ at small values
of separation parameter ($|\alpha _{0}|\le 0.8$). The physical
interpretation of this behaviour can be given as follows. The
contribution of vacuum (single-photon) state is dominant for the
even (odd) coherent state with small separation parameter.
According to the entropic analysis of quantum noise \cite{Mira00},
the Wehrl PD for vacuum is smaller than that for single-photon
state for all phases $\theta$.
%%%%%%%%%%%%%%%%%%%%%%%%%%%%%%%%%%%%%%%%%%%%%%%%%%%%%%%%%%%%%%%%%%%%%%%%%%%%
% figure 3.
\linebreak
\begin{figure}
\vspace*{0cm} \hspace*{2.1cm} \epsfxsize=7cm \epsfbox{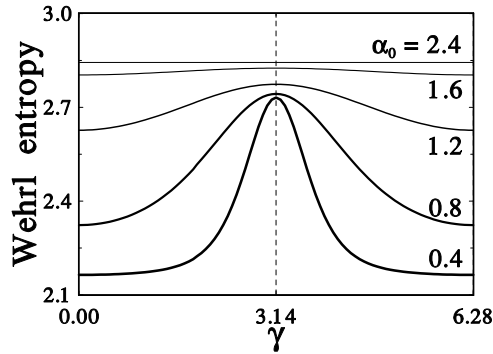}
\vspace*{0mm} \caption{ Wehrl entropy for Schr\"odinger cat
$|\alpha_0,\gamma\rangle$ in its dependence on the superposition
phase $\gamma$ for different values of the coherent amplitude
$\alpha_0$. }\label{fig3} \end{figure}%
\noindent This implies that $S^{\rm even}_\theta<S^{\rm
odd}_\theta$ for $|\alpha _{0}|\ll 1$. In contrast, this
inequality does not hold for the corresponding Husimi PDs: $P^{\rm
even}_\theta$ can be less but also greater than $P^{\rm
odd}_\theta$ for some values of phase $\theta$ at $|\alpha
_{0}|\le 0.8$. The Wehrl PD for the Yurke-Stoler coherent state,
$S^{\rm YS}_\theta$, approaches $S^{\rm odd}_\theta$ for
$\theta<\pi$ and $S^{\rm even}_\theta$ for $\theta>\pi$, as the
best presented in figure 2 for $|\alpha _{0}|=1.2$. In comparison,
the Husimi PD $P^{\rm YS}_\theta$ differs more significantly from
$P^{\rm even}_\theta$ and $P^{\rm odd}_\theta$ than the
corresponding Wehrl PDs for the same $|\alpha _{0}|<2.4$.

The Wehrl entropy for the cat, given by  (\ref{N07}), has been
studied numerically by Bu\v{z}ek et al.~\cite{Buze95b} for
arbitrary superposition phase $\gamma$.  Whereas, Jex and
Or\l{}owski~\cite{Jex94} and Vaccaro and Or\l{}owski~\cite{Vacc95}
studied the Wehrl entropy for the Yurke-Stoler coherent state
generated in a Kerr-like medium. In figure 3, we show the Wehrl
entropies $S_{\rm w}$ for the cat $|\alpha _{0},\gamma \rangle$ in
their dependence on the superposition phase $\gamma$ for various
values of the separation amplitude $\alpha_0$.  The curve for
$\alpha_0=0.8$ corresponds to the case analyzed by Bu\v{z}ek et
al.~\cite{Buze95b}. The discrepancies between Wehrl entropies for
the even, odd and Yurke-Stoler cats vanish with increasing
$|\alpha_0|$.  The Wehrl entropy in the high-amplitude limit
($|\alpha_0|\gg 1$) can be approximated by
%----------------------------------------------------------------------
\begin{eqnarray}
S_{\rm w}\approx 1+\ln \pi -\left| c_{1}\right| ^{2}\ln \left|
c_{1}\right| ^{2}-\left| c_{2}\right| ^{2}\ln \left| c_{2}\right|
^{2} \label{N19}
\end{eqnarray}
which for superpositions of equal-amplitude states reduces to
$S_{\rm w}\approx 1+\ln(2\pi)$. This value can be obtained by
integrating the approximate Wehrl PD, given by  (\ref{N15}). On
the scale of figure 3, the curve representing the Wehrl entropy
for $\alpha_0=2.4$ is practically indistinguishable from the
entropy in the infinite-amplitude limit.

The Schr\"{o}dinger cat-like state, also referred to as the kitten
state\footnote[1]{The term ``Schr\"{o}dinger's kitten'' in the
above meaning was coined by Agarwal et al.~\cite{Shan94}. In
contrast, some authors (e.g., Taubes \cite{Taub96}) prefer to use
this term to refer to a small (mesoscopic) Schr\"odinger cat,
which should be macroscopic according to original Schr\"odinger's
idea \cite{Schr35}.}, is a generalization of the Schr\"{o}dinger
cat for macroscopically distinct superposition state with more
than two components. In particular the normalized superposition
of $N$ coherent states:
%---------------------------------------------------------------------------
\begin{eqnarray}
|\alpha_0\rangle_N &=&\sum_{k=1}^{N}c_{k}|\exp ({\rm i}\phi
_{k})\alpha _{0}\rangle \label{N20}
\end{eqnarray}
is the standard example the Schr\"{o}dinger cat for $N=2$ and
Schr\"{o}dinger kitten for $N>2$. Equation (\ref{N20}) is valid
for arbitrary number, amplitudes and phases of the states in the
superposition. The state, defined by (\ref{N07}), is a special
case of  (\ref{N20}) for two coherent states with opposite phases
($\phi _{2}-\phi _{1}=\pi $) and $c_2/c_1=\exp({\rm i}\gamma)$.
The Husimi $Q$-function for the Schr\"{o}dinger cat-like state
reads as~\cite{Mira90,Buze95cat}:
%---------------------------------------------------------------------------
\begin{equation}
\hspace{-7mm} Q(\alpha )=Q_{0}(\alpha )\,+\,Q_{{\rm int}}(\alpha )
=\sum_{k=1}^{N}|c_{k}|^{2}\;Q_{k}(\alpha
)+2\sum_{k>l}|c_{k}||c_{l}|\;Q_{kl}(\alpha ) \label{N21}
\end{equation}
where the free part, $Q_{0}(\alpha)$, is the sum of the coherent
terms
%---------------------------------------------------------------------------
\begin{eqnarray}
Q_{k}(\alpha )&=&\frac{1}{\pi }\exp \left\{ -\left| \alpha -{\rm
e}^{{\rm i}\phi _{k}}\alpha _{0}\right| ^{2}\right\} \label{N22}
\end{eqnarray}
and the interference part, $Q_{\rm int}(\alpha )$, is given in
terms of
%---------------------------------------------------------------------------
\begin{eqnarray}
\hspace{-7mm} Q_{kl}(\alpha )&=&\sqrt{Q_{k}Q_{l}}\cos \Big[\gamma
_{k}-\gamma _{l} +2|\alpha |\;|\alpha _{0}|\cos (\phi
_{kl}^{(+)}+\theta _{0}-\theta )\sin \phi _{kl}^{(-)}\Big].
\label{N23}
\end{eqnarray}
The phases in  (\ref{N23}) are defined as $\gamma _{k}={\rm
Arg}\,c_{k}$, $\theta ={\rm Arg}\,\alpha $, $\theta _{0}={\rm
Arg}\,\alpha _{0}$, and $\phi _{kl}^{(\pm )}=\frac{1}{2}(\phi
_{k}\pm \phi _{l})$, where $\phi _{k}$ appears in  (\ref{N20}).
The Husimi $Q$-function, given by  (\ref{N21}), is a
generalization of  (\ref{N12}) for arbitrary number of components
in the superposition state. The compact-form exact analytical
expressions exist neither for the phase distributions nor the
Wehrl entropy. However, for well-separated (i.e., if
$|\alpha_0|\gg N$) states, the Wehrl PD for the Schr\"{o}dinger
cat-like state, given by  (\ref{N20}), is approximately equal to
%---------------------------------------------------------------------------
\begin{eqnarray}
S_{\theta }&\approx& \sum\limits_{k=1}^{N}|c_{k}|^{2} S^{\rm
cs}_{\theta }({\rm e}^{{\rm i}\phi _{k}}\alpha _{0})
-\sum\limits_{k=1}^{N}\left| c_{k}\right| ^{2}\ln (\left|
c_{k}\right| ^{2})P^{\rm cs}_{\theta }({\rm e}^{{\rm i}\phi
_{k}}\alpha _{0}). \label{N24}
\end{eqnarray}
The Wehrl PDs for the equientropic high-intensity states, given by
(\ref{N04}), were calculated from (\ref{N24}) and presented in
figures 1(d)--(f) in comparison to their Husimi
$Q$-representations given in figures 1(a)--(c), respectively. The
Wehrl PDs clearly show the number, amplitude and phase-space
configuration of the Schr\"{o}dinger
%%%%%%%%%%%%%%%%%%%%%%%%%%%%%%%%%%%%%%%%%%%%%%%%%%%%%%%%%%%%%%%%%%%%%%%%%%%%
% figure 4.
\linebreak
\begin{figure}
\vspace*{3mm} \hspace*{2.5cm} \epsfxsize=10.5cm \epsfbox{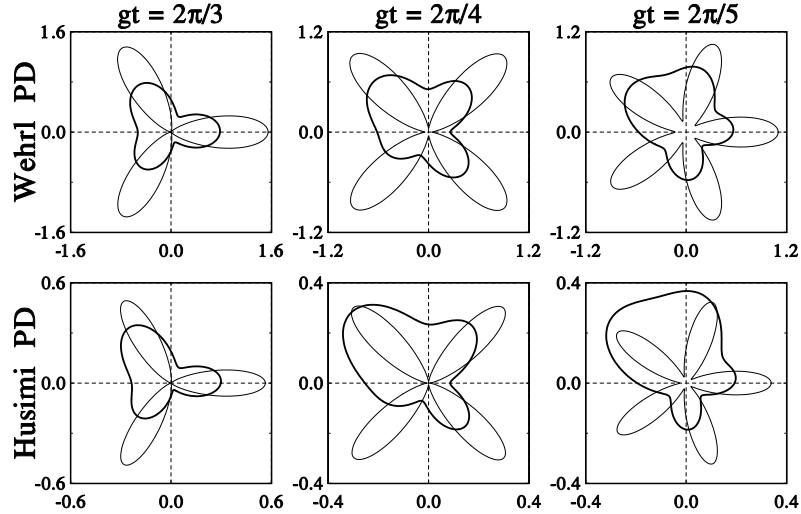}
\caption{ Polar plots of Wehrl and Husimi phase distributions for
the same Schr\"odinger cat-like states for Schr\"odinger cat-like
states generated in Kerr medium at different evolution times
$gt=2\pi/N$ ($N=3,4,5$) for initial coherent states with
amplitudes: $\alpha_0=3$ (thin curves) and $\alpha_0=\sqrt{2}$
(thick curves).}\label{fig4}
\end{figure}%
\noindent cat and cat-like states. We emphasize that the curves in
figures 1(d)--(f) cover approximately the same area equal to
$S_{\rm w}\approx 1+\ln 2\pi$. Eq. (\ref{N24}) for the
equal-amplitude superposition goes over into
%---------------------------------------------------------------------------
\begin{eqnarray}
S_{\theta }&\approx& \frac{1}{N}\sum_{k=1}^{N}\Big\{S^{\rm
cs}_{\theta }({\rm e}^{{\rm i}\phi _{k}}\alpha _{0})+P^{\rm
cs}_{\theta }({\rm e}^{{\rm i}\phi _{k}}\alpha _{0})\ln N\Big\}
\label{N25}
\end{eqnarray}
where $S^{\rm cs}_{\theta }(\exp \{{\rm i}\phi _{k}\}\alpha _{0},
0)$ are the coherent-field Wehrl PDs given by  (\ref{N16}), and
$P^{\rm cs}_{\theta }(\exp \{{\rm i}\phi _{k}\}\alpha _{0},0)$ are
the coherent-field Husimi PDs described by  (\ref{N17}). In
(\ref{N25}), we have assumed that the superposition coefficients
are the same for all components of the cat-like state, i.e.,
$c_{k}={\rm const}=1/\sqrt{N}$. Equation (\ref{N24}) leads, after
integration over $\theta$, to the Wehrl entropy
%---------------------------------------------------------------------------
\begin{eqnarray}
S_{\rm w}\approx 1+\ln \pi -\sum\limits_{k=1}^{N}\left| c_{k}\right|
^{2}\ln \left| c_{k}\right| ^{2}. \label{N26}
\end{eqnarray}
In the special case of equally-weighted superposition states,
(\ref{N26}) simplifies to $S_{\rm w}\approx 1+\ln (N\pi )$ in
agreement with the result of Jex and Or\l{}owski~\cite{Jex94}.
Estimations of the maximum number, $N_{\rm max}$, of
well-separated states in the superposition (\ref{N20}) as a
function of amplitude $|\alpha _{0}|$ can be given, e.g., by
\cite{Mira90}
%----------------------------------------------------------------------
\begin{equation}
N_{\max}(\alpha _{0}) = {\rm Int}(2^{-1/2}\pi|\alpha_0|)
\label{N27}
\end{equation}
where Int$(x)$ is the integer part of x. Approximations
(\ref{N24})--(\ref{N26}) are valid for $N\le N_{\rm max}(\alpha
_{0})$.

Several methods have been proposed to generate Schr\"odinger's cat
or cat-like states (see Refs.~\cite{Monr96,Frie00},
and~\cite{Buze95cat} for a review). In particular it has been
predicted that a coherent light propagating in a Kerr-like medium,
described by the Hamiltonian
%----------------------------------------------------------------------
\begin{eqnarray}
\widehat{H} &=& -\frac{1} {2}\hbar g\widehat{a}^{\dagger 2}
\widehat{a}^2 \label{N28}
\end{eqnarray}
where $g$ is the coupling constant, can be transformed into the
Schr\"odinger cat (Yurke-Stoler coherent state) \cite{Yurk86} and
kittens~\cite{Mira90,Aver89} at some evolution times. Explicitly,
for $gt=2\pi\frac{M}{N}$, where $M$ and $N$ are mutually prime
numbers, the generated state is a superposition of $N$ coherent
states, given by  (\ref{N20}), with the phases
$\phi_{k}=(2k+N-3)\frac{\pi}{N}$ and superposition coefficients
%----------------------------------------------------------------------
\begin{eqnarray}
c_{k}=\frac{1}{N}\sum_{n=1}^{N}\exp \left\{ {\rm i}n
\left[\frac{M}{N}\pi(n-1)-\phi_{k}\right]\right\}. \label{N29}
\end{eqnarray}
Jex and Or\l{}owski \cite{Jex94} and Vaccaro and
Or\l{}owski~\cite{Vacc95} have shown, by analyzing the model
described by (\ref{N28}), that the Wehrl entropy gives a clear
signature of the formation of the Schr\"odinger cat-like states.
The Wehrl PD, in comparison to the Wehrl entropy, offers more
detailed description of superposition states, showing explicitly
the phase configuration and amplitudes of the components. The
Wehrl and Husimi PDs are presented in figure 4 for the
Schr\"odinger cat-like states generated by Hamiltonian
(\ref{N28}) for different evolution times $gt$ in two regimes
determined by the initial amplitudes: $N<N_{\max}(3)=6$ and $N\ge
N_{\max}(\sqrt{2})=3$, where $N=3, 4, 5$. The number $N$ of
well-separated peaks in $S_\theta$ and $P_\theta$ clearly
corresponds to the number of states in the superposition. The
analysis of the Wehrl PD shows how the amplitude of the incident
coherent beam determines the maximum number of
well-distinguishable states. For example, both the Husimi and
Wehrl PDs depicted by thin lines in figure 4 have regular and
well-distinguishable structures even for five-component
superposition of the initial amplitude $|\alpha_0|=3$. However,
the four and five-component superpositions for the initial
condition $|\alpha_0|=\sqrt{2}$ are highly deformed as plotted by
the thick lines in figure 4. Thus, the Wehrl and Husimi PDs
describe the influence of the interference terms (\ref{N23}) on
formation of the Schr\"odinger cat-like states. Distributions
$P_\theta$ and $S_\theta$ are similar. However, a closer analysis
reveals their differences. In particular, as seen in figure 4 for
well-separated states, the rosette leaves are relatively broader
for $S_\theta$ than for $P_\theta$. For states not separated
distinctly, $S_\theta$ exhibits slightly more regular behaviour
than that of $P_\theta$. Nevertheless, the main difference and
advantage of our description in terms of $S_\theta$ over that of
$P_\theta$, resides in the area covered by $S_\theta$, which is
equal to the Wehrl entropy. Thus a simple phase-space operational
interpretation can be applied \cite{Buze95a}.

%%%%%%%%%%%%%%%%%%%%%%%%%%%%%%%%%%%%%%%%%%%%%%%%%%%%%%%%%%%%%%%%%%%%%%%%%%%%
\section{Conclusions}

The purpose of the paper was to find a good information-theoretic
measure of the superposition principle. We have shown that the
conventional Wehrl entropy is, in general,  {\em not} a good
measure for discriminating  two-component (Schr\"{o}dinger cats)
from multi-component (Schr\"{o}dinger kittens) macroscopically
distinct superposition states. We have applied a new information
measure -- the Wehrl phase distribution, which is a phase density
of the Wehrl entropy \cite{Mira00}. Compact form estimations of
both Wehrl measures were found for the superpositions of
well-separated states. It was demonstrated that the Wehrl phase
distribution, in contrast to the Wehrl entropy, properly
distinguishes the number and phase-space configuration of the
Schr\"{o}dinger cat and cat-like states even with
unequally-weighted components.

%%%%%%%%%%%%%%%%%%%%%%%%%%%%%%%%%%%%%%%%%%%%%%%%%%%%%%%%%%%%%%%%%%%%%%%%%%%%
\section*{Acknowledgments}

We thank \c{S}ahin K. \"Ozdemir, Masato Koashi, Yu-Xi Liu for
their stimulating discussions. JB was supported by the Czech
Ministry of Education (Grants No. LN00A015 and CEZ J14/98) and the
Grant Agency of Czech Republic (202/00/0142). MRBW acknowledges
the support from the Malaysia S\&T IRPA (Grant No. 09-02-03-0337).

%%%%%%%%%%%%%%%%%%%%%%%%%%%%%%%%%%%%%%%%%%%%%%%%%%%%%%%%%%%%%%%%%%%%%%%%%%%%

\vspace{5mm}
%%%%%%%%%%%%%%%%%%%%%%%%%%%%%%%%%%%%%%%%%%%%%%%%%%%%%%%%%%%%%%%%%%%%%%%%%%%%
{\setlength{\fboxsep}{10pt}
\begin{center}
\framebox{\parbox{0.75\columnwidth}{%
\begin{center}
Published in\\
J. of Physics A: Mathematical and General\\
{\bf 34} (2001) 3887-3896.
\end{center}}}
\end{center}

\end{document}